\documentclass[12pt,preprint]{aastex}

\newcommand\5{{\footnotesize V}}
\newcommand\4{{\footnotesize IV}}
\newcommand\3{{\footnotesize III}}
\newcommand\2{{\footnotesize II}}
\newcommand\1{{\footnotesize I}}
\newcommand\lam{$\lambda$}

\newcommand{\ea}{{et al.}}
\newcommand{\eg}{{\em e.g.}}

\newcommand\kms{\ensuremath{\mbox{km s}^{-1}}}

\newcommand{\p}{\ensuremath{\phantom{:}}}
\newcommand{\pp}{\ensuremath{\phantom{-}}}

\newcommand\Teff{$T_{\rm eff}$}

\slugcomment{Accepted by PASP}
\shorttitle{UV Spectra of B Stars in the SMC}
\shortauthors{Evans et al.}

\begin{document}


\title{The ultraviolet and optical spectra of luminous
B-type stars in the Small Magellanic Cloud}


\author{ C. J. Evans\altaffilmark{1},
         D. J. Lennon\altaffilmark{1},
         N. R. Walborn\altaffilmark{2},
         C. Trundle\altaffilmark{1,3}, 
         S. A. Rix\altaffilmark{1}
}

\altaffiltext{1}{Isaac Newton Group of Telescopes, 
                 Apartado de Correos 321, 
                 38700 Santa Cruz de la Palma, 
                 Canary Islands, 
                 Spain}
\altaffiltext{2}{Space Telescope Science Institute,
                 3700 San Martin Drive,
                 Baltimore,
                 MD 21218,
                 USA}
\altaffiltext{3}{Dept. of Pure and Applied Physics, 
                 The Queen's University of Belfast, 
                 Belfast, 
                 BT7 1NN,  
                 N. Ireland}

\begin{abstract}
We present ultraviolet spectra from the Space Telescope Imaging
Spectrograph (STIS) of 12 early B-type stars in the Small Magellanic
Cloud (SMC), comprising 9 supergiants and 3 giants.  A morphological
comparison with Galactic analogues is made using archival data from
the {\it International Ultraviolet Explorer (IUE)}. 
In general, the intensity of the P Cygni emission in the UV resonance
lines is greater, and seen to later spectral types, in the Galactic
spectra than in their metal-deficient SMC counterparts; we attribute these effects as
most likely arising from weaker stellar winds in the
SMC targets, as predicted by radiatively driven wind theory.

We also include unpublished STIS observations of two late O-type stars
in the SMC.  In combination with the B-type observations, and published O-type 
data, we now have an extensive ultraviolet spectral library of
metal-deficient stars, of use in the study of unresolved starbursts
and high redshift, star-forming galaxies.  In this context, we present
empirical measurements for the B-type spectra of the new `1978 index'
suggested by Rix et al. as a probe of metallicity in such systems.

\end{abstract}


\section{Introduction}
\label{intro}

Ultraviolet (UV) spectra are useful in the study of relatively nearby
blue galaxies undergoing intense star formation
(so-called starburst galaxies), and in high redshift star-forming
galaxies in which the stellar UV features are shifted into the
optical.  Studies of the composite spectra from such unresolved,
star-forming systems generally rely on spectral libraries to
investigate the stellar populations therein, with the libraries
compiled from UV observations of individual early-type
stars (e.g. Leitherer et al. 2001).

Distant star-forming systems are often less chemically evolved than
the Milky Way (e.g. Pettini et al. 2002).  With a metallicity of
roughly one-fifth solar (e.g.~Venn, 1999), the Small Magellanic Cloud
(SMC) offers an excellent opportunity for compilation of UV spectral
libraries more appropriate to the accurate interpretation of
metal-poor unresolved systems than if Galactic templates were used.
Libraries of high-quality UV data also enable detailed comparisons
with the synthetic spectra calculated from stellar atmosphere codes,
with a view to including theoretical spectra into population synthesis
models.

Walborn et al (2000; hereafter Paper~1) presented a high-resolution UV
spectral atlas of O-type stars in the SMC using data obtained with the
{\it Hubble\/ Space\/ Telescope (HST)} Space Telescope Imaging
Spectrograph (STIS) from Guest Observer program GO7437.  The
incorporation of these spectra into the population synthesis code {\sc
starburst}{\footnotesize 99}, and an assessment of their impact on the
resultant integrated spectra has been presented by Leitherer et
al. (2001).  However, as De Mello et al. (2000) demonstrated, the
contribution from B-type stars in integrated UV spectra can be
important in young stellar populations.  With this in mind we obtained
new STIS spectroscopy of 12 B-type stars in the SMC, with the sample
selected to cover a range of effective temperatures and luminosities.
These data have already been used to constrain the age of the stellar
population observed in a `super-star cluster' in the starburst galaxy
NGC 1705 (V$\acute{\rm a}$zquez et al. 2004), illustrating their
usefulness in the study of unresolved populations.

The new B-type STIS spectra have a much greater wavelength
coverage than those in Paper~1 (\lam\lam1150-2350
cf. \lam\lam1150-1750), achieved by observing each target
using the echelle gratings at two central wavelengths (see Section
\ref{obsdetails}).  Our motivation behind this strategy was to obtain
high resolution observations of the iron `forest' of absorption lines
from Fe$^{2+}$, which lies in the \lam\lam1850-2100 region.  Since
these observations, Rix et al. (2004) have published a theoretical
study in which synthetic UV spectra from the WM-Basic code (Pauldrach
et al. 2001) are included in {\sc starburst}{\footnotesize 99}.  They
advocate an equivalent-width index centered at 1978 \AA\/ as a useful
metallicity indicator, and the new STIS data presented here offer an
opportunity to begin to calibrate this index empirically.

Aside from their use as spectral templates, UV observations of
individual early-type stars in the SMC enable detailed studies of
stellar evolution and radiatively driven winds in a metal-poor
environment.  For instance, some of the O-type spectra presented in
Paper~1 were used in detailed model atmosphere analyses by Bouret et
al. (2003) and Hillier et al. (2003).  Similarly, four of the B-type
STIS spectra discussed here were used by Evans et al. (2004a) to
determine stellar wind terminal velocities ($v_\infty$), that were 
then used in the analysis of the optical spectra by Trundle et al. (2004).

Here we primarily concern ourselves with a discussion of the 
morphological phenomena seen in the UV spectra of
our SMC targets, and compare them with Galactic analogues observed
with the {\it International\/ Ultraviolet\/ Explorer (IUE)} telescope.
We also discuss the correlation between the new UV data and
high-resolution optical data of our targets, in particular including
observations of the H$\alpha$ Balmer line.

\section{Observations}
\label{obsdetails}

\subsection{UV spectroscopy}

For our $HST$ General Observer program (GO9116) we observed
12 B-type stars in the SMC using the STIS; observational parameters of
our targets are given in Table~\ref{targets}.  Each star was observed
at two central wavelengths in echelle mode; with the far-UV MAMA
(Multi-Anode Microchannel Array) detector and the E140M grating
(centered at 1425 \AA), and with the near-UV MAMA and E230M grating
(centered at 1978 \AA).  The echelle observations were made through
the $0{\farcs}2~\times~0{\farcs}2$ entrance aperture and the effective
spectral resolving powers of the E140M and E230M gratings are $R =
46,000$ and 30,000 respectively.  The exposure times were set
from consideration of low-resolution $IUE$ spectra and are given in
Table \ref{targets}.

The primary {\sc calstis} processing steps are described in Paper~1,
with the difference here that the two-dimensional inter-order
background correction performed manually in those reductions has now
been incorporated into the standard pipeline as {\sc sc}{\footnotesize
2}{\sc dcorr}.  As described by Evans et al. (2004a), after pipeline
reduction the individual echelle orders were extracted and merged to
form a continuous spectrum.  Prior to merger (typically) 25 pixels at
both ends of each order are clipped; the spectral overlap between
orders is generally large enough to accommodate this and the final
result has a more consistent signal-to-noise ratio than that obtained
otherwise.  Blueward of 1650 \AA\/ there is a significant decrease in
the signal-to-noise ratio of the E230M data. therefore this region is
removed.  The E140M and E230M spectra are then merged to form a final
spectrum covering 1150-2350 \AA.  For display purposes (and for
consistency with Paper~1) the spectra are binned to 0.25\,\AA~per data
point.  The signal-to-noise ratio (on the basis of photon statistics)
is typically in the range of 20-30 per binned data point.

In Table \ref{targets} we also give H~\1 column densities, $N$(H~\1),
toward our targets, found assuming that Lyman-$\alpha$ has a pure
damping profile (\eg~Shull \& Van Steenberg 1985).  To avoid possible
contamination by the N~\5 \lam\lam1239, 1243 doublet, we consider only
the blueward wing when deriving the tabulated values.  The column
densities are consistent with those for other stars in the SMC,
e.g. Tumlinson et al. (2002).

There are archival spectra from the {\it Far Ultravoiolet
Spectroscopic Explorer (FUSE)} of the type Ia stars, however many
suffer from large column densities of H$_2$ and, in the context of the
present discussion, reveal little further information such that they
are not considered here (A. Fullerton, private communication, 2004).

\subsubsection{Additional UV observations of O-type stars}
\label{7437}

Observations for the GO7437 program were incomplete when Paper 1
was written.  We now briefly discuss additional STIS 
observations with the E140M grating of AzV\,423 (Azzopardi \& Vigneau
1975, 1982) and NGC~346 MPG~487 (Massey et al. 1989).  The optical
spectrum of AzV~423 was classified as O9.5 II(n) by Walborn et
al. (2002), in their discussion of its spectrum from $FUSE$; the $HST$
data for this star have not been published previously.  NGC~346
MPG~487 was classified as O8 V by Evans (2001) and Bouret et al.
(2003), with the latter presenting a model atmosphere analysis
in which parts of the STIS spectrum were shown (their Figure~9)
These two stars were not incorporated in {\sc starburst}{\footnotesize 99}
with the main O-type sample (Leitherer et al., 2001) and offer
additional early-type template spectra for population synthesis work.

\begin{figure*}
\begin{center}
\includegraphics{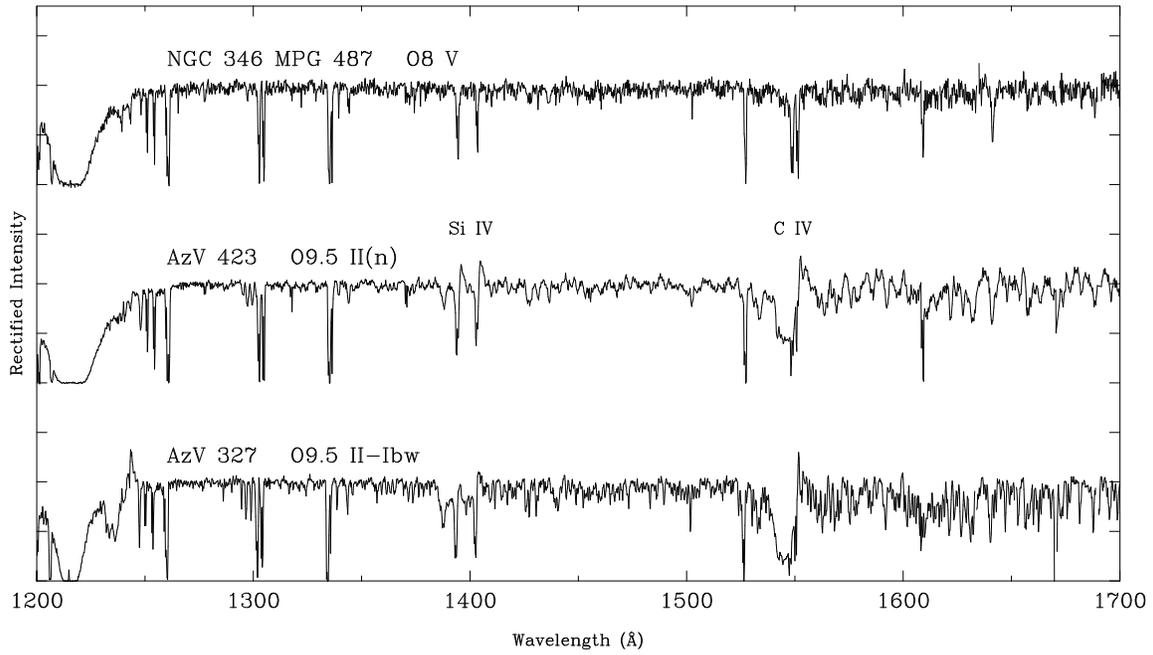}
\caption{STIS spectra of AzV~423 and NGC~346 MPG~487 from {\it HST} program GO7437.  
The spectrum of AzV~327 was published previously by Walborn et al. (2000) and is 
included here for comparison with AzV~423 (see discussion in Section \ref{7437}).}
\label{extras}
\end{center}
\end{figure*}

The two STIS spectra are shown in Figure~\ref{extras}.  
The spectrum of NGC~346 MPG~487 is primarily photospheric in origin
and complements the data in Paper~1 well, providing better sampling of
the main sequence.  The new spectrum of AzV~423 is more interesting,
particularly in comparison to that of AzV~327 from Paper~1 (also shown 
in Figure~\ref{extras}; the Si~\4
and C~\4 profiles are more strongly in emission in AzV~423, although
the C~\4 absorption is weaker.  In contrast, the N~\5 doublet displays
relatively strong emission in AzV~327, compared to none in AzV~423.
We also have unpublished, red optical data for these two stars from
the University College London echelle spectrograph (UCLES) at the 3.9-m
Anglo-Australian Telescope (AAT); the H$\alpha$ Balmer profiles of
both stars are in absorption and are practically identical.
Comparisons of the blue optical data (see Paper~1 and Walborn et al. 2002)
are complicated somewhat by the slightly broader lines in AzV~423, but
it appears that C~\3 is stronger, and N~\3 weaker, in AzV~423.  We
suggest that the differences in the UV spectra may arise from
different amounts of mixing of CNO-processed material into the
atmosphere and winds of the two stars.  Note that the spectrum of
AzV~327 in Figure~\ref{extras} appears to be noisier than that of
AzV~423; the exposure times were identical and the cosmetic difference
arises from the (apparently) faster rotation of AzV~423.
Considering their similar magnitudes (suggesting similar physical
luminosities) future quantitative analysis of these stars would
be of interest.

\subsection{Optical data}
In addition to the UV data we also consider high-resolution optical
spectra of our targets.  Blue and red region optical spectra of the
eight targets from the AzV catalogue were obtained with the UV-Visual
Echelle Spectrograph (UVES) at the ESO Very Large Telescope (VLT) in
November 2001.  The effective resolving power of the UVES set-up was
$R \sim 20000$ and the signal-to-noise ratio of the data was in excess
of 100.  Full observational details are given by Trundle et
al. (2004).  For AzV\,22 and 362, the H$\alpha$ spectra are those from
the New Technology Telescope (NTT) using the ESO Multi Mode Instrument
(EMMI).  These are preferred over the VLT data following the
discussion of nebular contamination by Trundle et al.

Blue region spectra for the four targets in NGC\,330 were obtained
using the CASPEC spectrograph at the ESO 3.6-m telescope in August 1994
and September 1996.  The resolution of the spectra is comparable to the UVES
data and the signal-to-noise ratio ranges from 40 to 90; further details
are given by Lennon et al. (2003).

\section{Morphological comparison with Galactic $IUE$ spectra}
\label{morph}

\subsection{Luminosity class Ia stars}
\begin{figure*}
\begin{center}
\includegraphics[width=22cm, height=16cm, angle=-270]{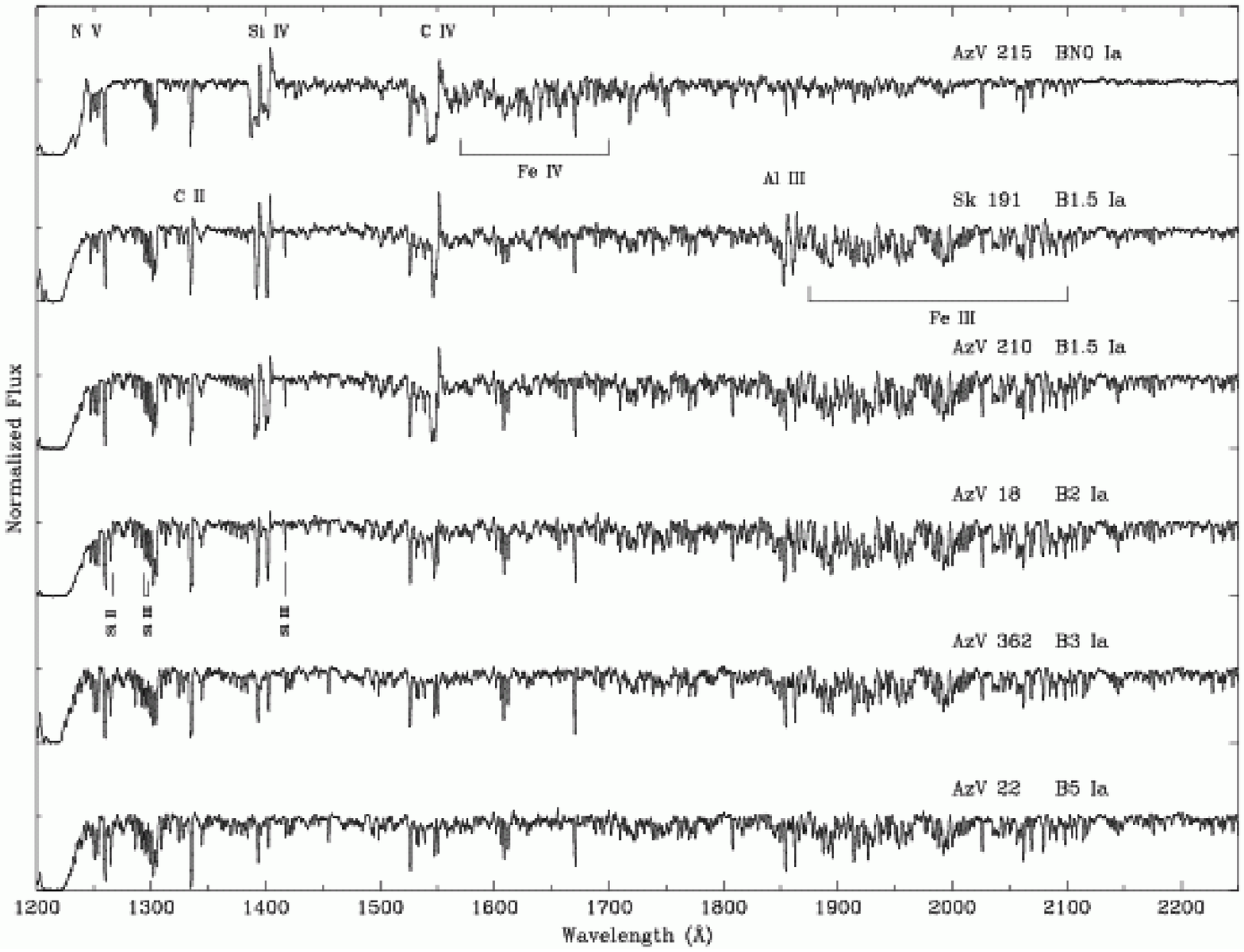}
\caption{\footnotesize STIS spectra of class Ia SMC stars.  The lines 
identified in AzV~215 are, from left to right, N~\5 \lam\lam1239,
1243; Si~\4 \lam\lam1394, 1403; and C~\4 \lam\lam1548, 1551.
Additional lines marked in Sk~191 are, C~\2 \lam\lam1334, 1336; and
Al~\3 \lam\lam1855, 1863; and in AzV~18, Si~\2 \lam1265 and Si~\3
\lam\lam1294, 1299, 1417.  Also note the iron `forests' from Fe~\4 in
AzV~215 (\lam1550-1750) and Fe~\3 (\lam1850-2100) in the other
spectra.  Each spectrum is offset by two continuum flux units.}
\label{fig1}
\end{center}
\end{figure*}

\begin{figure*}
\begin{center}
\includegraphics[width=15.7cm, height=16cm, angle=-270]{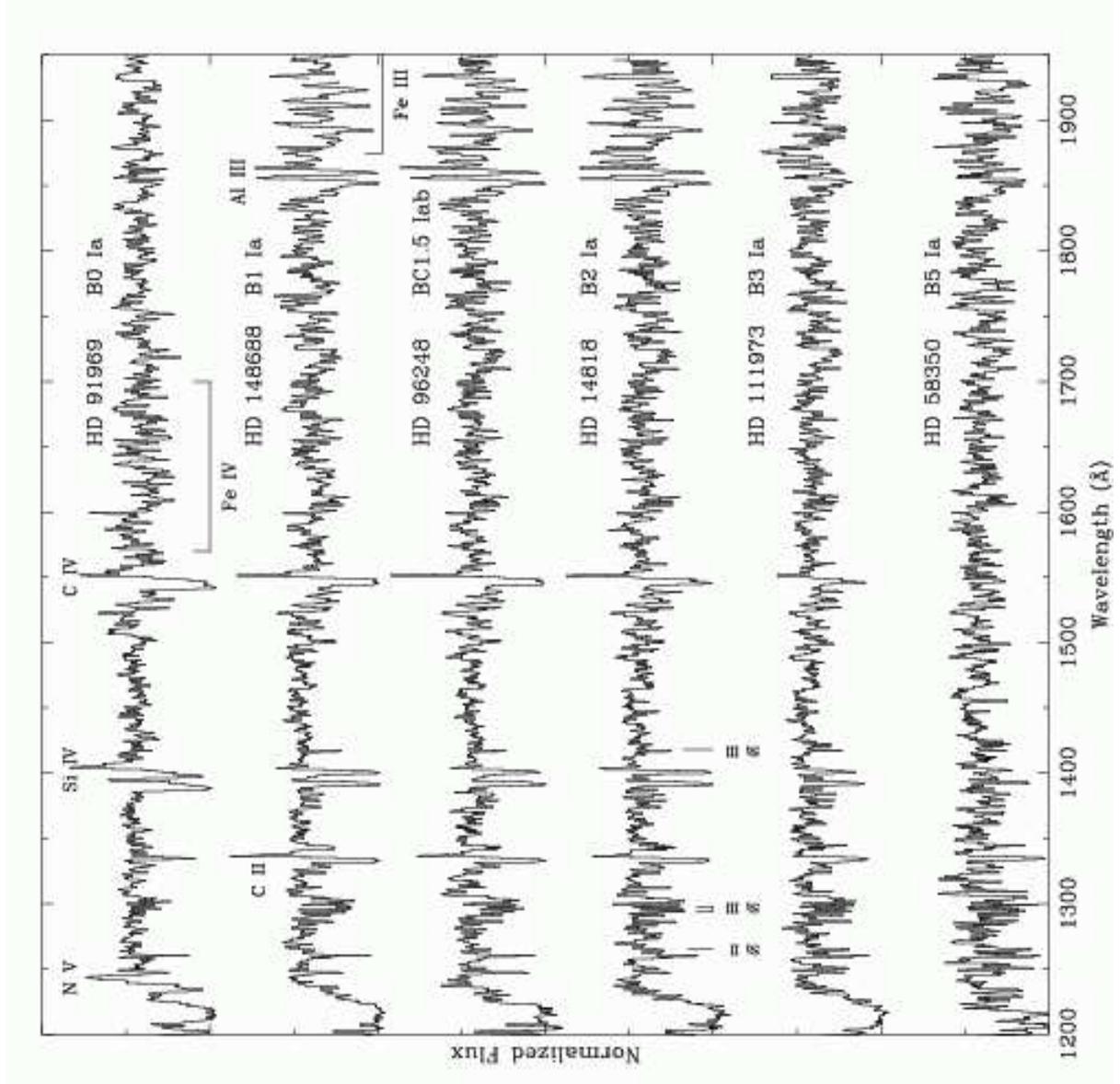}
\caption{$IUE$ spectra of Galactic analogues to the SMC stars shown in 
Figure \ref{fig1}; the line identifications are the same.}
\label{fig2}
\end{center}
\end{figure*}

The STIS spectra of our class Ia SMC targets (with the exception of
AzV~104, see Section \ref{optical}) are shown in Figure \ref{fig1}.  Galactic
analogues to these six stars from the Walborn et al. (1995a) $IUE$
atlas are shown in Figure \ref{fig2}.  For display purposes the $IUE$
spectra have been smoothed using a 5-pixel `boxcar' median
filter. There are extant high-resolution, long-wavelength camera
observations (i.e., redward of 1950\,\AA) in the $IUE$ archive for the
majority of the stars here, however the spectra are typically of 
low quality and are not included in our discussions.  When compiling
Galactic analogues we have deliberately omitted HD\,152236
($\zeta^1$~Sco, B1.5 Ia$+$), because of its peculiar nature, and identification as
a candidate luminous blue variable (e.g., Sterken et al. 1997).
Instead we include HD\,148688 (B1~Ia) and HD\,96248 (BC1.5~Iab) as the
most suitable stars for comparison with Sk~191 and AzV~210 (both
B1.5~Ia).  As noted by Evans et al. (2004a; Fig.~2 therein), the morphology of the
Si~\4 doublet in AzV~215 is peculiar; there is a relatively flat
`shelf' in the profiles, resembling those seen in $\kappa$~Ori
(HD~38771, B0.5 Ia, Walborn et al. 1995a), the origins of which are
currently unclear.

As one progresses from B0 to B5, there are a number of clear trends.
In the SMC data the P Cygni profiles of the Si~\4 and C~\4 lines
weaken at later types (see Figure \ref{fig1}); there
is evidence of a weak stellar wind in the Si~\4 lines of AzV~18
(B2~Ia), with only photospheric features seen in AzV~362 (B3 Ia).
The $HST$ Faint Object Spectrograph (FOS) observations of AzV~488 
(B0.5~Ia) presented by Walborn et al. (1995b) fit neatly into this
sequence.  The SMC data contrast with the Galactic data (see Figure \ref{fig2}),
in which obvious wind signatures are seen through to at least
B3~Ia (HD\,111973).  

Furthermore, at a given spectral type, the P Cygni emission
in the SMC targets is generally weaker than in the Galactic
data.  This is a well documented phenomenon in early-type spectra
(e.g., Hutchings, 1982; Bruhweiler et al. 1982) and is most striking
in the current comparisons for C~\2 \lam\lam1334, 1336 and Al~\3
\lam\lam1855, 1863.  In contrast, the Si~\4 \lam\lam1394, 1403 lines 
in HD~96248 and 148688 are interesting in that, although they display
weak emission, the intensity is less than in the SMC spectra.  In the
case of HD~96248 this could be potentially explained as arising from
its lower luminosity, with the significant C~\2 and C~\4 emission due
to an enhanced carbon abundance (the star is classified as BC-type).
However, the Al~\3 features in HD~96248 are comparable to those seen
in HD~148688, suggesting that abundance anomalies are not solely
responsible.

It is apparent from Figures \ref{fig1} and \ref{fig2} that the lower
metallicity of the SMC is manifested by significantly weaker Fe~\3
absorption in the `forest' redward of the Al~\3 doublet, although the
extent of the $IUE$ data in this region is relatively limited.  Figure
\ref{fig1} also neatly illustrates the change of the dominant
ionization stage of iron in early B-type stars.  In AzV~215 the
multitude of Fe~\4 lines in the \lam1550-1700 region are clearly
visible, with relatively weak Fe~\3 lines; this situation is reversed
in Sk~191 and AzV~210.  These features are also present in the
Galactic data, although it is less obvious given the greater
metallicity and reduced wavelength coverage.  Walborn \&
Nichols-Bohlin (1987) noted that there appeared to be a distinct
difference in the stellar wind features between B0.5~Ia and B0.7~Ia
spectra.  This correlates with the change in iron ionization seen 
here, suggested by Vink et al. (2001) as a physical explanation
for the so-called `bi-stability' of stellar winds.

\subsection{Lower luminosity stars}
\begin{figure*}
\begin{center}
\includegraphics[width=22cm, height=16cm, angle=-270]{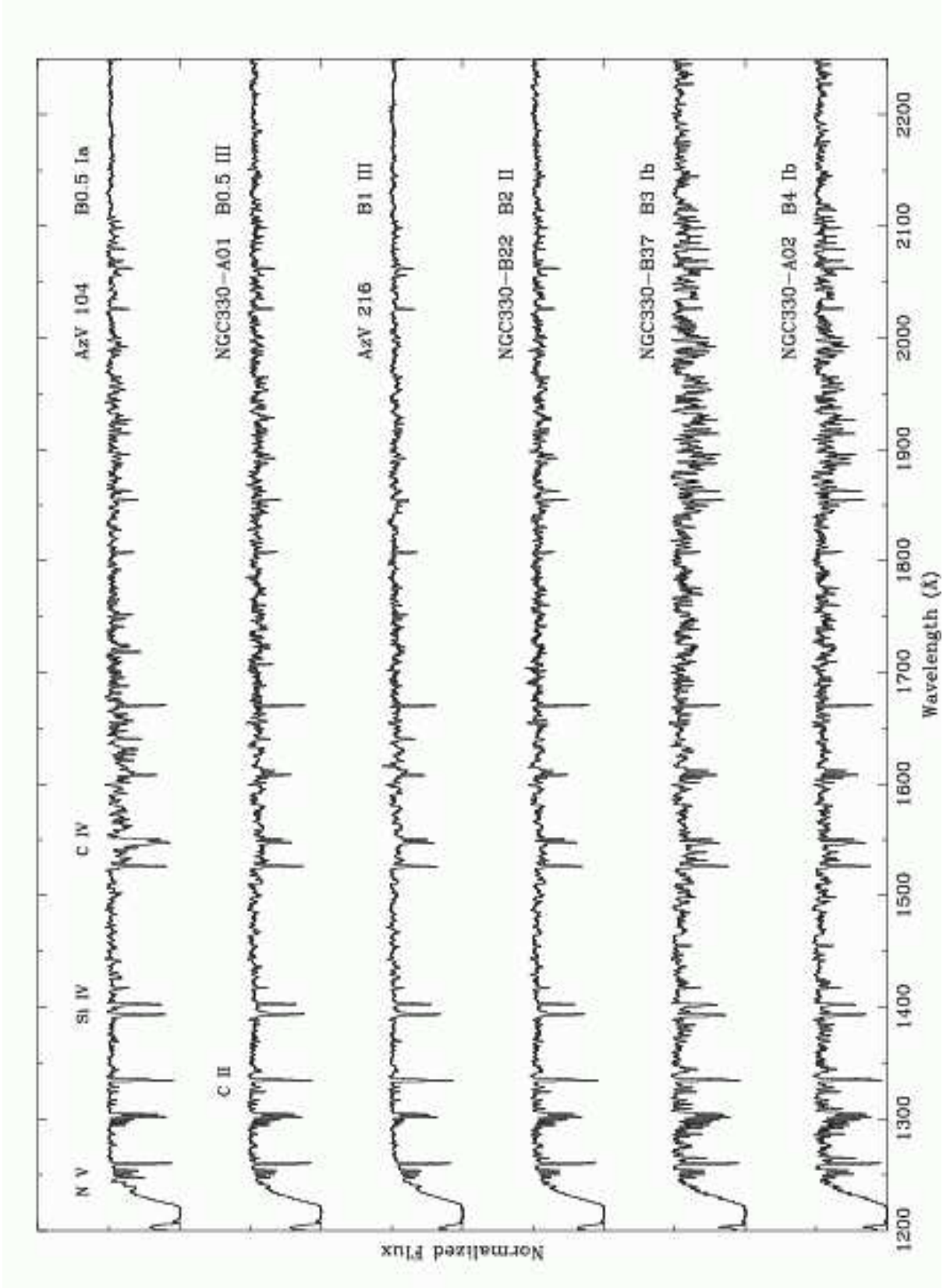}
\caption{STIS spectra of additional B-type SMC targets.}
\label{fig3}
\end{center}
\end{figure*}

\begin{figure*}
\begin{center}
\includegraphics[width=15.7cm, height=16cm, angle=-270]{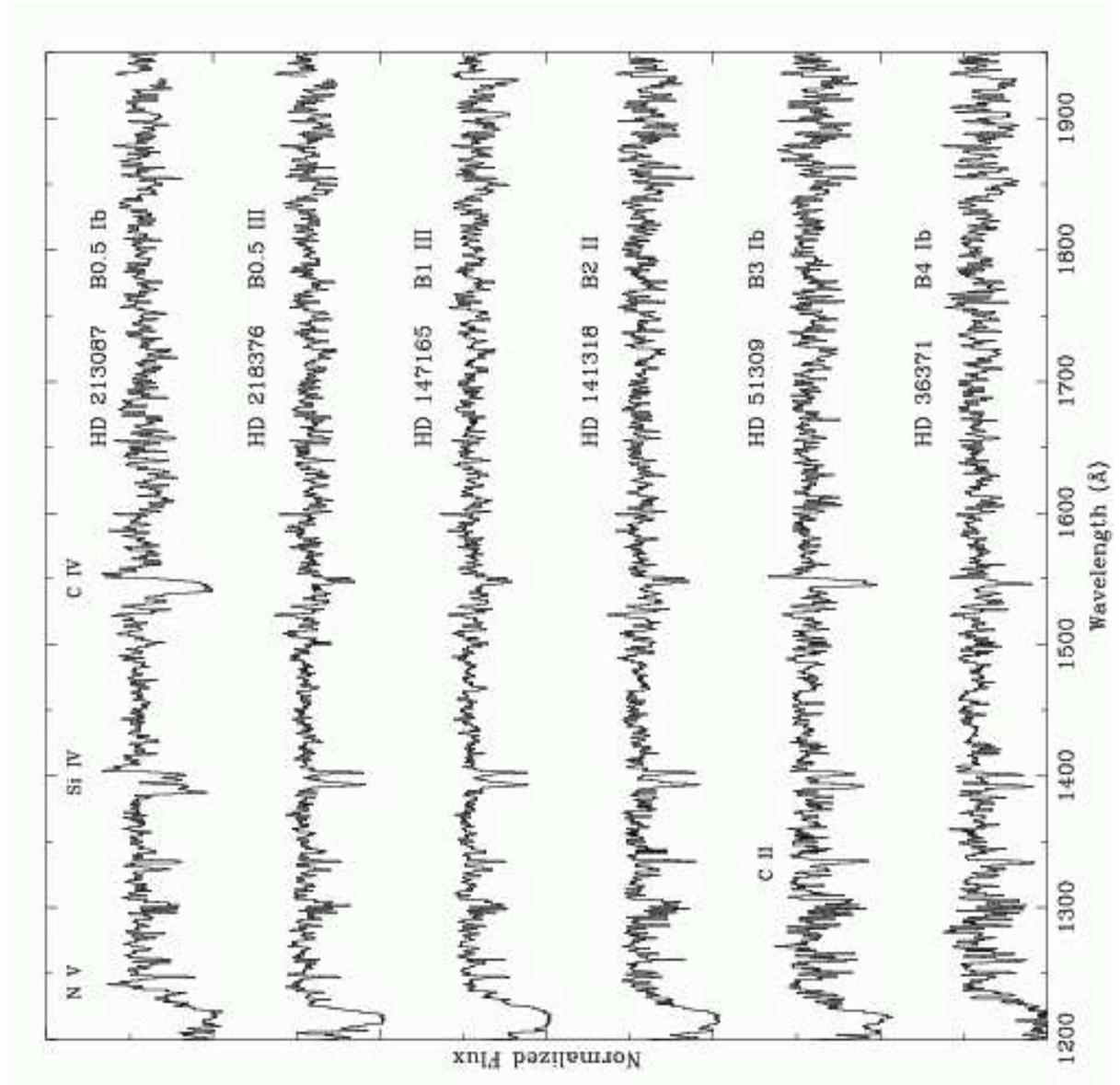}
\caption{$IUE$ spectra of Galactic analogues to 
those in Figure \ref{fig3}; the luminosity type of HD~213087 is
deliberately mis-matched to that of AzV~104 (see text for discussion).}
\label{fig4}
\end{center}
\end{figure*}

We now move to the less luminous $HST$ targets, 
shown in Figure \ref{fig3}.  In Figure \ref{fig4} we again show Galactic
analogues drawn from the atlas of Walborn et al. (1995a).

Classified as B0.5 Ia by Lennon (1997), AzV~104 appears to be
exceptional in that its physical properties do not appear to correlate
well with its blue-region optical morphology.  A critical
re-examination of the optical spectra of this star is presented in
Section \ref{optical}.  For now we compare our STIS spectrum with that
of HD\,213087 (B0.5~Ib).  In their analysis of terminal velocities,
Evans et al. (2004a) were unable to apply the `Sobolev with exact integration'
(SEI) method to the UV resonance lines of AzV~104,
although from inspection of the Si~\4 and C~\4 profiles it appears
that there is some wind present, albeit particularly weak.  In
contrast, the wind in HD~213087 appears more significant (ignoring for
the moment any potential luminousity effects arising from mis-matching
classifications).

The remainder of the lower luminosity STIS spectra are largely
photospheric in origin, with no evidence of stellar winds in their
resonance lines.  Similarly, in the Galactic spectra, HD~218376
(B0.5~III), HD~147165 (B1~III), and HD~141381 (B2~II) also contain
only photospheric features.  However, in the two class Ib Galactic
spectra, i.e., HD~51309 and 36371, weak P Cygni profiles are seen in
the C~\4 doublet.  As in the class Ia stars, evidence of stellar winds
is seen to later spectral types for class Ib in the Galaxy than in the
SMC.  Again, the dominant stage of iron is revealed by the spectra,
the \lam\lam1550-1700 Fe~\4 lines are clearly visible in AzV~104, with the
\lam\lam1850-2000 Fe~\3 lines strongly in absorption in NGC330-B37 and
NGC330-A02.

\section{Morphological comparisons of the ultraviolet and optical data}
\label{optical}
\begin{figure*}
%\begin{center}
\includegraphics[width=22cm, height=16cm, angle=-270]{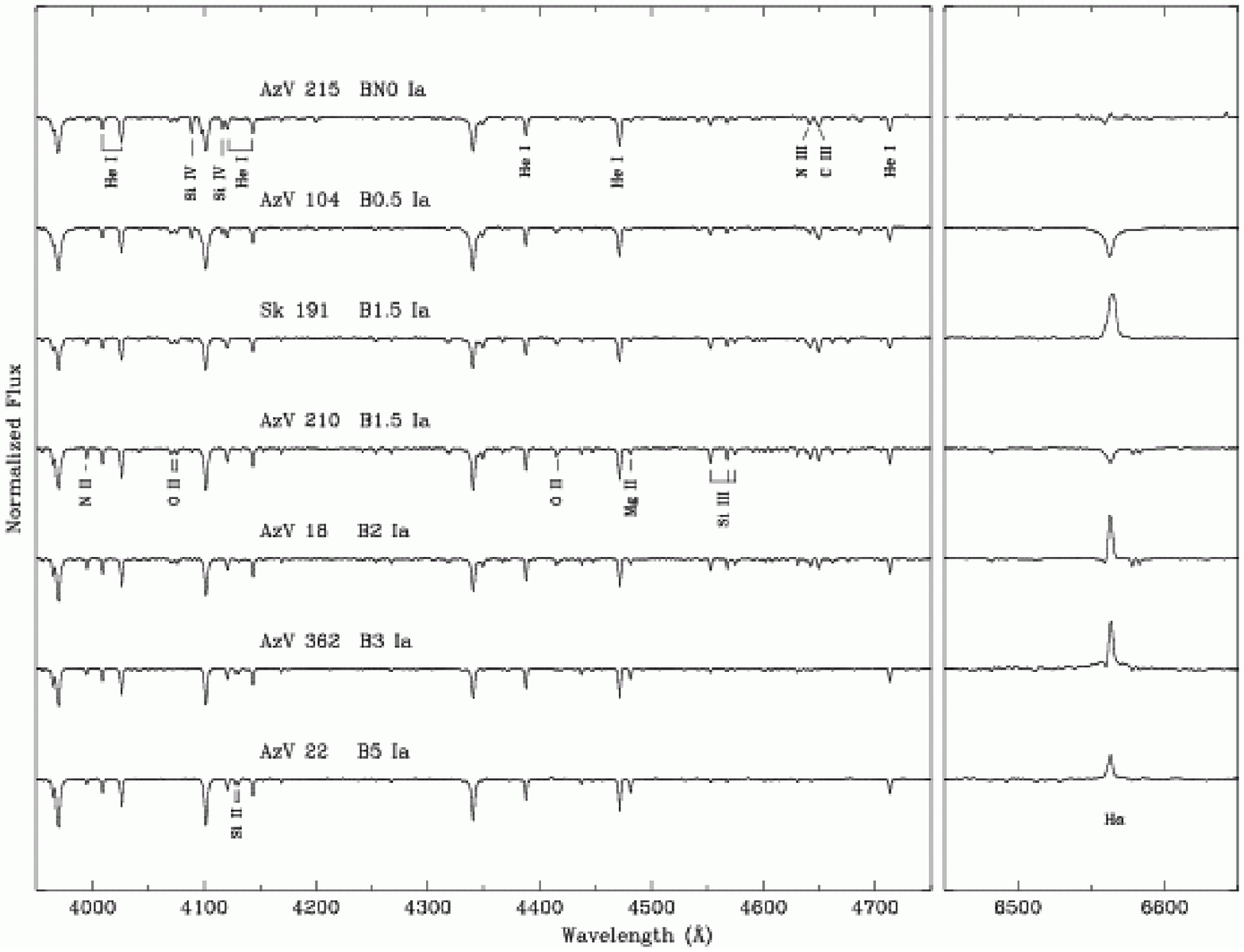}
\caption{\footnotesize Optical spectra of class Ia SMC stars.  
Lines identified in the AzV~215 spectrum are, from left to right by
species, He~\1 \lam\lam4009, 4026, 4121, 4144, 4388, 4471, 4713; and
the N~\3 \lam4640 and C~\3 \lam4650 blends.  Lines identified in
AzV~210 are N~\2 \lam3995; O~\2 \lam\lam4072-4476, 4415-4417; Mg~\2
\lam4481; and Si~\3 \lam\lam4553, 4568, 4575.  Lines
marked in AzV~22 are Si~\2 \lam4128-32; and the Balmer H$\alpha$ line
(\lam6563).  For display purposes the data have been smoothed to an
effective resolution of 1.0 \AA~and each spectrum is offset by one
continuum flux unit.}
\label{opt}
\end{figure*}

In Figure \ref{opt} we show optical spectra for the seven class Ia stars.
The data have been smoothed and rebinned to an effective resolution of
1.5~\AA\/ (cf. Walborn \& Fitzpatrick, 1990) and have also been
corrected for the measured recession velocities.

Given their excellent signal-to-noise ratio, the blue-region
optical data offer a wonderfully clear illustration of the trends in
early B-type spectra.  The principal temperature diagnostic in these
stars is the ionization balance of Si~\4, Si~\3, and Si~\2.  The
optical lines marked in Figure \ref{opt} are well complemented by
several UV photospheric lines  (e.g., Si~\2 \lam1265, and Si~\3
\lam\lam1294, 1299, 1417, as marked in Figure \ref{fig1}) that were 
suggested by Massa (1989) and Prinja (1990) as potential UV
temperature diagnostics.  Indeed, from simple inspection of the Si~\2
\lam1265 line alone in AzV~210, one would be tempted to suggest a cooler
temperature than Sk~191, even though the spectral types are identical.
These morphological arguments are confirmed by the model atmosphere
analysis of Trundle et al. (2004) which found a temperature of 20,500~K
for AzV~210, some 2,000~K lower than that found for Sk~191.

Similarly, the optical spectra neatly illustrate the increase and then
decrease in the intensity of the main O~\2 features (\lam\lam4072-76,
4345-49, 4415-17) as one moves from
B0$\rightarrow$B1.5$\rightarrow$B3.  The H$\alpha$ Balmer profiles of
the stars are also shown in Figure \ref{opt}, with a wide range of
morphologies displayed, i.e., pure absorption (albeit very slightly
`filled-in' by wind effects, AzV\,104), `filled-in' absorption
(AzV\,210), weak P-Cygni (AzV\,215) and broad emission (Sk\,191).

From comparisons with the UV spectra in Figure~\ref{fig1} (and AzV\,104
in Figure~\ref{fig3}), the optical data highlight the strong
correlation of the UV and H$\alpha$ profiles for the earliest B-type
spectra, i.e., AzV\,215, AzV\,104, AzV\,210 and Sk\,191.  In contrast, the
significant H$\alpha$ emission in AzV~18 (B2~Ia), AzV~362 (B3~Ia), and
AzV~22 (B5~Ia) appear initially to contradict the UV spectra; the
H$\alpha$ emission suggests some degree of stellar wind, yet the UV
spectra display little or no evidence.  This is attributable to the
range of temperatures covered by our sample, e.g., the dominant
ionization stage of silicon changes from Si$^{3+}$ $\rightarrow$
Si$^{2+}$ $\rightarrow$ Si$^{1+}$ as one progresses from B0 to B5.
Thus, even in the presence of a stellar wind, P Cygni emission is not
expected in the Si~\4 doublet at the latest types, although weak
photospheric features are still seen.

\subsection{AzV~104}
Particularly striking in Figures \ref{fig4} and \ref{opt} are the
spectra of AzV\,104 (B0.5 Ia), in which the H$\alpha$ absorption tallies with
the lack of obvious wind signatures in the UV; in this case it would
appear that the luminosity has not been predicted correctly by
morphological considerations.  
The optical classification employed here is that from Lennon
(1997), in which luminosity class was assigned using the H$\gamma$
equivalent width calibrations of Azzopardi (1987).  From photometric
and intrinsic colour arguments, Trundle et al. (2004) gave $M_v =
-$5.82, more in keeping with that expected from a B0.5 Ib star (e.g.,
Humphreys \& McElroy, 1984) than one classified as B0.5 Ia.  

The overestimated luminosity is not obviously a metallicity effect.
The principle advantage of using H$\gamma$ as a luminosity diagnostic
is that it is expected to be relatively unaffected by changes in metallicity.
In contrast, if Galactic criteria involving metal-line intensities are 
applied directly to B-type SMC spectra, the luminosity class is often
underpredicted (e.g. Walborn 1983). 
We do not suggest that the luminosity class be
revised on the basis of this discussion (such a change would against the
philosophy of the `MK process'), but merely comment that in this case 
the H$\gamma$ calibrations appear to incorrectly predict the intrinsic
luminosity.

\section{Discussion}
\label{discussion}

\subsection{Weaker winds at low metallicity?}
A long-standing challenge of observational studies of massive stars
has been to test conclusively the theoretical prediction that
mass-loss scales with metallicity such as $\dot{M}(Z) \propto
Z^{0.5\rightarrow0.7}$ (Kudritzki et al. 1987; Vink et al. 2001).

Recent results from Evans et al. (2004b) and Trundle et al. (2004)
have highlighted that the predictions from the mass-loss `recipe' of
Vink et al. (2001) do not tally with the observationally derived rates
in the Magellanic Clouds.  However, contemporary results from Crowther
et al. (2004) suggest similar differences between observational and
theoretical mass-loss rates for Galactic B-type supergiants, i.e.,
there appears to be tentative evidence for observational confirmation
of some metallicity dependence, with the absolute values differing to
those of the Vink et al. (2001) purely as a consequence of the
theoretical methods employed.

The comparison of the Galactic and SMC spectra in Section \ref{morph}
offers some insight in this context.  In general, from the more
intense P Cygni emission at a given spectral type, a stronger stellar
wind can be inferred in the Galactic stars than in our SMC targets.
At first, such behaviour in the Clouds was attributed solely to the
lower abundances (e.g., Bruhweiler et al. 1982), with wind effects
thought to play a less significant role.  However, in a star such as
AzV~215, the resonance lines are strongly saturated and the line is
only weakly dependent on abundance changes, even over the large range
of SMC to solar metallicity; the stellar wind dominates the formation
of the resulting profile.

Stronger winds in the Galactic targets can also be inferred from the
presence of P Cygni features at later spectral types than those seen
in the SMC spectra.  This point is made most strongly by the saturated
C~\4 doublet in HD~51309 (B3~Ib, see Figure \ref{fig4}) compared to
the weak photospheric absorption seen in NGC330-B37 (also B3~Ib); such
a difference cannot be explained by abundance effects alone.

It is unlikely that different temperatures can account for these
effects; the results found by Trundle et al. (2004) for their SMC
targets are comparable to those found for Galactic B-type supergiants
by Crowther et al. (2004).  Similarly, though a relatively wide range
of physical luminosity is possible for stars classified as type Ia,
the Galactic and SMC samples do not appear to be particularly
mismatched.  For instance, Kudritzki et al. (1999) give the luminosity
of HD~14818 (B2 Ia) as log(L/L$_\odot$) = 5.47, cf. 5.44 for AzV~18
(also B2 Ia, Trundle et al. 2004).

Interestingly, Evans et al. (2004a) found no evidence for a significant
scaling in the wind terminal velocities in early B-type UV spectra;
four of their targets were drawn from the STIS data presented here
(namely AzV\,18, AzV\,210, AzV\,215, and Sk~191).  So, in comparison to Galactic
stars, the early B-type SMC supergiants appear to have weaker winds,
but with comparable terminal velocities, in agreement with the predictions
of radiatively driven wind theory.

\subsection{Ultraviolet Iron Indices and their application to unresolved systems}
\label{Fe_index}

The STIS data have a further application in the context of population
synthesis models.  Rix et al. (2004) have proposed an equivalent-width
index centered at 1978~\AA\/ as a useful probe of metallicity in
high redshift, star-forming galaxies.  This index is effectively the
summed photospheric absorption between 1935 and 2020
\AA\/ (which largely arises from Fe~\3 lines, e.g., Swings et
al. 1976) and, in the case of constant star formation, was found to be
more sensitive than the indices suggested by Leitherer et al. (2001)
centered at 1370~\AA\/ and 1425~\AA.

The data presented here provide an opportunity to begin to
investigate this index empirically.  Following correction for the
stellar radial velocities given in Table \ref{targets}, each of the
STIS spectra was treated in the same manner as the models used by Rix
et al., i.e., they were degraded to an effective resolution of 2.5
\AA\/ (to ensure compatibility with contemporary observations of high 
redshift systems) and then rectified using the same continuum windows.  We
include the equivalent widths of the 1978 index in Table
\ref{targets}.  The measurement uncertainties largely
arise from the difficulty in accurate placement of the continuum and
are typically $\pm$10$\%$ when W$_\lambda$(1978) $>$ 5 \AA, increasing
to 15-20$\%$ when W$_\lambda$(1978) $<$ 5 \AA.  The measurements for
the class Ia stars provide a further illustration of the temperature
sequence in early B-type spectra.  The equivalent width is greatest at
B2, and is seen to increase significantly between B0.5 and B1.5, the
region in which the `bi-stability jump' in the behaviour of stellar
winds is found.

As one might expect, there is some degeneracy between the index for
the early-B, class Ia objects and later-type, less luminous stars
(i.e., AzV\,215 and AzV\,104 cf. NGC330-B22); similar conclusions were
reported by Swings et al. (1976) from their low-resolution spectra,
obtained using the European TD1 satellite.  In a brief episode of star
formation this degeneracy (effectively the age/metallicity degeneracy)
can be quite limiting but, coupled with other information from the
spectrum such as the appearance of the resonance lines, the dominant
stellar population can be better constrained.  For example, the
dominant population in an integrated spectrum displaying significant P
Cygni emission in the resonance lines, with a large 1978 equivalent
width, is likely to be much less evolved than one in which there is no
evidence of wind features, coupled with a smaller 1978 index.  Similar
morphological arguments were used by V${\rm \acute{a}}$zquez et
al. (2004), in combination with theoretical evolutionary isochrones to
constrain the age of the super-star cluster NGC~1705-1.

\subsection{The combined SMC sample}

\begin{figure*}
\begin{center}
\includegraphics[scale=1.0, angle=0]{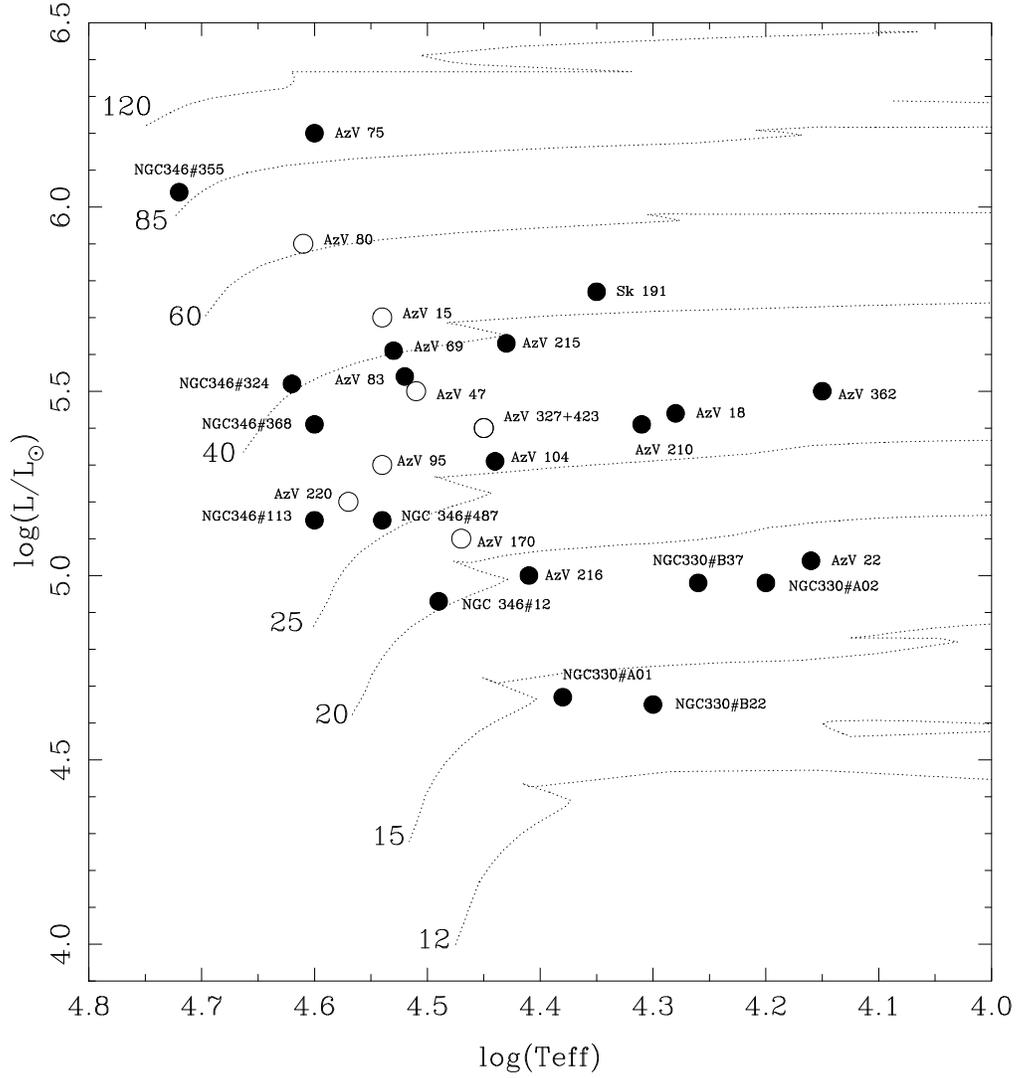}
\caption{H-R diagram of the full {\it HST} STIS sample.  Evolutionary tracks ({\it dotted lines})
are from Charbonnel et al. (1993) and are labelled with initial stellar masses
in $M_\odot$.  Stars marked with open symbols are those for which
physical parameters have been interpolated from published results (see
text for details).}
\label{hrd}
\end{center}
\end{figure*}

In Table \ref{hrdtab} we compile the full list of stars observed with
$HST$ for our O- and B-type programs (i.e., GO7437 and GO9116) and in
Figure \ref{hrd} we present an H-R diagram for the sample.  The
evolutionary tracks (dotted lines) are from Charbonnel et al. (1993),
with $Z$~=~0.004 and standard mass-loss rates; for clarity the 120,
85, 60 and 40 M$_\odot$ models are truncated at the onset of helium
burning.  Our intent here is merely to guide the reader to the
approximate mass range of our targets so for simplicity we prefer
these over newer evolutionary tracks (e.g. those of Maeder \& Meynet,
2001), in part to remove the need to assume initial rotational
velocities for each star.

Where available, temperatures and luminosities have been taken from
published model atmosphere analyses, as indicated in Table
\ref{hrdtab}.  For the remaining stars, temperatures have been interpolated
between recent results from line-blanketed model atmospheres (Martins
et al. 2002, Crowther et al. 2002, and Herrero et al. 2002), with
luminosities calculated from the photometry in Paper~1 and, in the
case of AzV~423, Azzopardi \& Vigneau (1975).  Intrinsic colours were
taken from Fitzgerald (1970) and, for simplicity, the bolometric
corrections were calculated using the relation given by Vacca et
al. (1996; their Eqn. 6).  Although their study employed atmospheres
that did not consider the effects of line blanketing, the lower
temperatures derived for O-type stars from its inclusion will
yield sensibly smaller bolometric corrections using their relation,
sufficiently accurate for our `cosmetic' H-R diagram.

\section{Summary}

We have compared our STIS observations of early B-type stars in the
SMC with Galactic spectra from the $IUE$ archive.  The two main
morphological features of note are the generally weaker P Cygni
emission profiles in the SMC spectra at given spectral type, and the
presence of P Cygni profiles in the Galactic spectra to later types
than those seen in the SMC.  The UV and optical spectra of our most
luminous SMC targets were compared, neatly illustrating the correlation
between diagnostic features in both regions.

We have also begun attempts to calibrate empirically the new 1978
index suggested by Rix et al. (2004).  Given the relative dearth of
high-resolution observations in this region, these data will also
enable valuable tests of WM-Basic (Pauldrach et al. 2001) or
{\sc cmfgen} (Hillier \& Miller 1998), necessary if model spectra
are to be routinely used in population synthesis models.

As demonstrated by Figure \ref{hrd}, the full SMC OB-type sample now
covers a significant part of the upper H-R diagram, providing an
unprecedented library of high-resolution, metal-poor spectra.  One
issue that we have not addressed at the current time is the wealth of
interstellar information in the combined OB-type sample; these data
provide a wide variety of sight-lines to the SMC and will form
the basis of a future study.

\section{Acknowledgements}
CJE (under grant PPA/G/S/2001/00131), DJL, and SAR acknowledge
financial support from the UK Particle Physics and Astronomy Research
Council ({\sc pparc}).  NRW acknowledges support through grant
GO-9116.09 from the STScI.  Based in part on observations with the
NASA/ESA {\it Hubble Space Telescope} obtained at the Space Telescope
Science Institute (STScI), which is operated by the Association of
Universities for Research in Astronomy, Inc., under NASA contract NAS
5-26555; and on INES data from the $IUE$ satellite.  We thank Paul
Crowther for providing his results in advance of publication, 
Linda Smith for the red AAT observations, Alex Fullerton for the
extracted $FUSE$ spectra, and the referee for their helpful comments.

%%%%-----TABLES-----

\begin{deluxetable}{lllcccccccccccc}
\rotate
\tabletypesize{\scriptsize}
\tablewidth{0pc}
\tablecolumns{15}
\tablecaption{Observational parameters of GO9116 target stars \label{targets}}
\tablehead{
\colhead{Star} & \colhead{Alias} & \colhead{Sp. Type} & \colhead{Ref.} & \colhead{$V$} & \colhead{$B-V$} & \colhead{Ref.} & \colhead{$M_{V}$} & 
\colhead{\Teff} & \colhead{$v_r$} & \multicolumn{2}{c}{Exp. [min]} & log $N$(H~{\tiny I}) & \colhead{Optical} & \colhead{W$_\lambda$(1978)}\\
 & & & & & & & & \colhead{[kK]} & \colhead{[\kms]} & \colhead{E140M} & \colhead{E230M} & \colhead{[cm$^{-2}$]} & \colhead{Telescope} & \colhead{[\AA]}
}
\startdata
AzV 215    & Sk 76  & BN0 Ia           & 2 & 12.69 & $-$0.09 & 4 & $-$6.6 & 27.0 & 159 & 106 & 37.5  &    21.8  & VLT       &   5.7 \\
AzV 104    & --     & B0.5 Ia          & 2 & 13.17 & $-$0.16 & 4 & $-$5.8 & 27.5 & 163 & 106 & 37.5  &    21.4  & VLT       &   5.0 \\
NGC330-A01 & --     & B0.5 III         & 3 & 14.70 & $-$0.18 & 3 & $-$4.5 & 24.0 & 151 & 237 &  145  &    21.5  & ESO 3.6-m &   4.2 \\
AzV 216    & --     & B1 III           & 5 & 14.22 & $-$0.13 & 4 & $-$5.1 & 26.0 & 203 & 183 &  108  &    21.6  & VLT       &   3.1 \\ 
Sk 191     & --     & B1.5 Ia          & 2 & 11.86 & $-$0.04 & 1 & $-$7.4 & 22.5 & 130 &  54 & 37.5  &    21.4  & VLT       &  15.1 \\
AzV 210    & Sk 73  & B1.5 Ia          & 2 & 12.60 & $-$0.02 & 4 & $-$6.7 & 20.5 & 173 & 124 & 60.5  &    21.6  & VLT       &  16.3 \\
AzV 18     & Sk 13  & B2 Ia            & 2 & 12.46 & \pp0.03 & 1 & $-$7.0 & 19.0 & 138 & 216 &   75  &    21.8  & VLT       &  18.5 \\
NGC330-B22 & --     & B2 II            & 3 & 14.29 & $-$0.13 & 3 & $-$4.9 & 20.0 & 157 & 308 &  243  &    21.5  & ESO 3.6-m &   4.7 \\
AzV 362    & Sk 114 & B3 Ia            & 2 & 11.36 & $-$0.03 & 1 & $-$7.8 & 14.0 & 208 &  54 & 37.5  &    21.3  & VLT/NTT   &  14.2 \\
NGC330-B37 & --     & B3 Ib            & 3 & 13.19 & $-$0.07 & 3 & $-$6.0 & 18.0 & 156 & 183 &  108  &  \p21.4: & ESO 3.6-m &  13.9 \\
NGC330-A02 & --     & B4 Ib            & 3 & 12.90 & $-$0.05 & 3 & $-$6.3 & 16.0 & 151 & 237 &  145  &    21.5  & ESO 3.6-m &   9.8 \\
AzV 22     & Sk 15  & B5 Ia            & 2 & 12.25 & $-$0.10 & 1 & $-$6.6 & 14.5 & 139 & 162 &   75  &    21.5  & VLT/NTT   &  11.3 \\
\enddata
\tablecomments{Stellar identifications are those of 
Azzopardi \& Vigneau (AzV; 1975, 1982), Robertson (NGC330; 1974) and Sanduleak (Sk; 1968). 
Absolute magnitudes ($M_v$) are calculated using intrinsic colours
from Fitzpatrick \& Garmany (1990) and Fitzgerald (1970), adopting a
distance modulus of 18.9 (Harries \ea, 2003) and taking the ratio of
total to selective extinction $R = 3.1$ (\eg~Schal$\acute{\rm e}$n,
1975).  Effective temperatures and radial velocities are from Lennon et al. (2003) and Trundle et al. (2004).}
\tablerefs{(1) Garmany et al. (1987); (2) Lennon (1997); (3) Lennon et al. (2003) -- although in the case of 
NGC330-A01 the luminosity class has been revised from III/V to III following comparisons of the H$\gamma$ 
equivalent width with the calibrations of Balona \& Crampton (1974); (4) Massey (2002); 
(5) Trundle et al. (2004)}
\end{deluxetable}

\clearpage

\begin{deluxetable}{llccc}
\tabletypesize{\scriptsize}
\tablewidth{0pc}
\tablecolumns{5}
\tablecaption{Physical parameters of the combined SMC sample\label{hrdtab}}
\tablehead{
\colhead{Star} & \colhead{Spectral Type} & \colhead{\Teff (kK)} & \colhead{log(L/L$_\odot$)} & \colhead{Ref.}
}
\startdata
NGC 346 MPG 355 &  ON2 III(f$^\ast$)& 52.5 & 6.04 & 1 \\
NGC 346 MPG 324 &  O4 V((f))        & 41.5 & 5.52 & 1 \\
NGC 346 MPG 368 &  O4-5 V((f))      & 40.0 & 5.41 & 1 \\
%NGC 346 MPG 435 &  O4-6 III(f)      &&&\\
AzV 80          &  O4-6n(f)p        & {\it 40.5} & {\it 5.9} & -- \\
%AzV 75          &  O5 III(f$+$)     & {\it 40.5} & {\it 6.1} & -- \\
AzV 75          &  O5 III(f$+$)     & 40.0 & 6.2  & 5 \\
NGC 346 MPG 113 &  OC6 Vz           & 40.0 & 5.15 & 1 \\
AzV 220         &  O6.5 f?p         & {\it 37.3} & {\it 5.2} & -- \\
AzV 15          &  O6.5 II(f)       & {\it 35.0} & {\it 5.7} & -- \\
AzV 95          &  O7 III((f))      & {\it 34.7} & {\it 5.3} & -- \\
AzV 83          &  O7 Iaf$+$        & 32.8 & 5.54 & 2 \\
AzV 69          &  OC7.5 III((f))   & 33.9 & 5.61 & 2 \\
AzV 47          &  O8 III((f))      & {\it 32.6} & {\it 5.5} & -- \\
NGC 346 MPG 487 &  O8 V             & 35.0 & 5.15 & 1 \\
%NGC 346 MPG 682 &  O9-9.5 V         &&&\\
AzV 327         &  O9.5 II-Ibw      & {\it 28.4} & {\it 5.4} & -- \\
AzV 423         &  O9.5 II(n)       & {\it 28.4} & {\it 5.4} & -- \\
NGC 346 MPG 12  &  O9.5-B0 V (N str)& 31.0 & 4.93 & 1 \\
AzV 170         &  O9.7 III         & {\it 30.0} & {\it 5.1} & --\\
AzV 215         &  BN0 Ia           & 27.0 & 5.63 & 4 \\
AzV 104         &  B0.5 Ia          & 27.5 & 5.31 & 4 \\
NGC330-A01      &  B0.5 III/V       & 24.0 & 4.67 & 3 \\
AzV 216         &  B1 III           & 26.0 & 5.00 & 4 \\
Sk 191          &  B1.5 Ia          & 22.5 & 5.77 & 4 \\
AzV 210         &  B1.5 Ia          & 20.5 & 5.41 & 4 \\
AzV 18          &  B2 Ia            & 19.0 & 5.44 & 4 \\
NGC330-B22      &  B2 II            & 20.0 & 4.65 & 3 \\
AzV 362         &  B3 Ia            & 14.0 & 5.50 & 4 \\
NGC330-B37      &  B3 Ib            & 18.0 & 4.98 & 3 \\
NGC330-A02      &  B4 Ib            & 16.0 & 4.98 & 3 \\
AzV 22          &  B5 Ia            & 14.5 & 5.04 & 4 \\
\enddata
\tablerefs{(1) Bouret et al. (2003); (2) Hillier et al. (2003); (3) Lennon et al. (2003); 
(4) Trundle et al. (2004); (5) Massey et al. (2004)}
\tablecomments{Italicized values denote values interpolated from published results, see
text for details}
\end{deluxetable}
\clearpage

%%%%-----FIGURES----

\clearpage

\clearpage

\clearpage

\clearpage

\clearpage

\clearpage

\end{document}